\documentclass[12pt]{article}
\usepackage{amsmath,amsfonts}

\begin{document}

\begin{center}
\textbf{Proof of the hypothesis Edmonds's, not polynomial of NPC-problems and classification of the problems with polynomial certificates}

B.S. Kochkarev

{Kazan (Volga region) federal university, Russia}
E-mail: bkochkar@kpfu.ru
\end{center}

\begin{abstract}
We show that the affirmation $P\subseteq NP$ (in computer science) erroneously and we prove the justice of the hypotesis J.Edmonds's $P\neq NP$. We show further that all the $NP$-complete problems is not polynomial and we give the classification of the problems with the polynomial certificates.
\end{abstract}

In 1964 Alan Kobham [1] and, independently, in 1965 Jack Edmonds [2] have entered a concept of complexity class $P$.

Definition 1 [1,2]. A language $L$ belong to $P$ if there is an algorithm $A$ that decide $L$ in polynomial time ($\leq O(n^{k})$) for a constant $k$. Class of problems $P$ is called polynomial.

According [3] J.Edmonds has entered also the complexity class $NP$. This is the class of the problems (langages) that can be verified by a polynomial-time algorithm.

Definition 2 [3]. A langage $L$ belongs to $NP$ if there exists a two-input polynomial-time algorithm $A$ and such polynomial $p(x)$ with whole coefficients that

$L=\{x\in \{0,1\}^{*}:$ there exists a certificate $y$ with $\mid y\mid \leq p(\mid x\mid)$ and $A(x,y)=1\}$.

 In this case we say that the algorithm $A$ verifies the language $L$ in polynomial time.

 According to definition 2, if $L\in P$ and $\mid y\mid \leq p(\mid x\mid)$, then $L\in NP$. But if $L\in P$ and length of the certificate not polynomial from length $x$, then $L\notin NP$.

  J. Edmonds has conjectured also that $P\neq NP$. In [4] we builded one class of the polynomial problems  with not polynomial certificates. According of the considerations see above immediately it follows that $P\neq NP$. In [3,5,6] authors consider only two version $P\subseteq NP$ ($P\subset NP, P=NP$) and the version of J. Edmonds $P\neq NP$ generally reject.

 In 1971 S.A.Cook has put the question: "whether can the verification of correctness of the decision of a problem be more long than the decision itself independently of algorithm of verification?" This problem have a relation to cryptography. In other formulation this problem look so: whether can to build a cipher such that his decipher algorithmically more complicated than find of cipher?

 In 2008 [7] we have proposed a model of decision Cook's problem: let $M$ and $M'$ are two sets such that $M$ is decide in polynomial time, let then there exists the injective map $\phi$ of $M$ in $M'$ such that for any $m\in M$ $\phi (m)$ find in not polynomial time. In [4,8,9] we have cited also some realizations of this model.

 Definition 3 [3]. A language $L\subseteq \{0,1\}^{*}$ is $NP-complete (NPC)$ if

 1. $L\in NP$, and

 2. $L'\leq _{P}L$ for every $L'\in NP$.

If a language $L$ satisfies property 2, but not necessarily property 1, we say that $L$ is $NP-hard$.

Theorem 1. Any $L\in NPC$ is not polynomial-time solvable.

Proof. Evidently, sufficiently it prove for a $NPC$ problem. Let $S=\{1,2,...,n\}$ is the set of the natural numbers, where $n$ is the enough large odd natural number. Let further $q(x)$ is the prime algebraic polynomial (with whole coefficients) of degree $\geq 5$. Let $j$ is a number of $S$. We form the family of the all subsets $F\subset S$ such that $j\notin F$ and $\mid F\mid =\frac {n-1}{2}$. Further we add to a subset $F$ number $j$ and we mark this subset by $\tilde{F}=F\bigcup \{j\}$. We consider the following problem: be in need of an urn containing the cards with the writing subsets $F$ draw out the card with $\tilde{F}$ after ${ n-1 \choose  {(n-1)}/{2}}$  of number of extractions. If in ${ n-1 \choose  {(n-1)}/{2}}$ step the card with $\tilde{F}$ not appear then the all extractions cards return in the urn and the process of extractions of the cards before appearance of the card with $\tilde{F}$ renew. Evidently, since ${ n-1 \choose  {(n-1)}/{2}}$ [11] is exponent, then our problem solvable in exponential-time. The certificate for this problem is the polynomial $q'(x)=(x-j)q(x)$. In order to verify whether appear the decision correct it is necessary calculate the meaning of the polynomial $q'(x)$ for $x=j$. If $q'(j)=0$, then the decision is correct. Thus the problem citing belongs to $NPC$ and not polynomial.

 In definition 2 of class $NP$ of J.Edmonds figure the notion of the certificate. Thus $L\in NP$ if the arguments of the checking algorithm such that a) for any admissible entrance of the length $n$ $\mid x\mid \leq O(n^{k})$, where $k$ is constant and $x\in L$ is the decision for the entrance correspondent; b) $y$ is the algebraically polynomial from $x$ (polynomial from $x$ with whole coefficients).

Definition 4. We will say that the certificate $y$ is polynomial if $y$ is the algebraically polynomial from $x$.

Since the principal notion in Definition 2 is the notion of certificate which form the checking algorithm, then in present work we give the classification of problems having the certificate $y=p(x)$, where $p(x)$- is the algebraically polynomial from $x$.

Theorem 2. Let $L$ be language with the polynomial certificate $y$. Then is true one from following versions:

a) $L\in P$,

b) $L$ is the algorithmically undecided language,

c) $L\in NPC$,

d) $L$ is language with the not trivial polynomial certificate $y$.

Proof. a)Obviously, the problem of sorting it is the polynomial problem [7], with certificate $y=x$, which belongs to $NP$.

b) Let $p(x)=0$ be an unsolvable algebraic equation. This equations was found for the first time by E.Galoua. As such concrete equation can give any the algebraic equation $p(x)=0$, where $p(x)$ the prime polynomial of degree $\geq 5$. Problem whether there exists the algorithmic decision of this equation: the answer is negative. In this case the certificate $y$ is the polynomial $p(x)$.

c) Let $L\in NPC$. Above we noted (theorem 1), that in this case $L$ is not polynomial. Evidently, all problems from $NPC$ have the trivial certificates $y=x$.

d) Let $n\geq 5$ be a natural number. Let further $2,3,...,p_{n}$ be first $n$ of the simple numbers.Evidently,

$q(x)= p_{j}x^{r}+p_{i}x^{r-1}+p_{i}x^{r-2}+...+p_{i}x+2\cdot 3\cdot 5\cdot ...\cdot p_{n-1}\cdot p_{n}$,  (1)

where $r\geq 5, p_{i}\neq p_{j}$ is the polynomial which according to criterion of Eisenstein it is the prime polynomial. We give the algebraic equation

$(x-p_{i})q(x)=0$  (2)

The equation (2) have the decision $x=p_{i}$. We consider the following problem: be in need of an urn containing the balls with the numbers $2,-2,3,$ $-3,...,p_{n},-p_{n}$ drawn out the ball with the number $p_{i}$. Evidently, this problem solvable in exponential-time by means of extraction of balls from the urn without of the return. In the very unreasonable case we draw out the ball with the number $p_{i}$ over $2n$ steps. The number of the ways of such issue is $(2n-1)!$, that is the exponent from $n$. The certificate for this problem is the polynomial $(x-p_{i})q(x)$. In order verify whether appear the decision correct it is necessary make of the factorisation of the free member of the polynomial $q(x)$. The number drawn ball is decision of problem, if it is among of factors of the free member of $q(x)$. We say, that the certificate of this problem (1) differ from certificates of the problems of c).

\centerline{REFERENCES}
\smallskip

1. A.Cobham, The intrinsic computational difficulty of functions // In Procedings of the 1964 Congress for Logic, Methodology, and the Philosophy of Science.- North-Holland, 1964.-P.24-30.

2. J.Edmonds, Paths, trees and flowers // Canadian Journal of Mathema\-tics.-1965-Vol.17.-P.449-467.

3. T.H.Cormen, Ch.E.Leiserson, R.L.Rivest, Introduction to Algorithms, MIT Press, 1990.

4. B.S.Kochkarev, About one class polynomial problems with not polynomial certificates, arXiv:1210.7591v1 [math.CO] 29 Oct 2012.

5. S.A.Cook, The $P$ versus $NP$ Problem, Manuscript prepared for the Clay Mathematics Institute for the Millennium, April, 2000, www.cs.toronto.edu sacook.

6. A.A.Razborov, Theoretical Computer Science:vzglyad mathematica,\\http://old.computerra.ru/offline/2001/379/6782/.

7. B.S.Kochkarev, On Cook's problem, http: //www.math.nsc.ru/conference/ malmeet/08/ Abstract/Kochkarev.pdf.

8. B.S.Kochkarev, Prilogenie monotonnykh funktsiy algebry logiki k probleme Kuka, Nauka v Vuzakh:matematika, fizika, informatika, Tezisy dokladov Mejdunarodnoj nauchno-obrazova\-tel\-noi konferentsii, 2009, pp. 274-275.

9. B.S.Kochkarev, K probleme Kuka, Matematicheskoje obrazovanije v shkole i v vuze v uslovijakh perekhoda na novye obrazovatelnye standarty, Materialy Vserossijskoj nauchno-practicheskoj konferentsii s mejdunarodnym uchastiem, Kazan, 2010, pp.133-136.

10. B.S.Kochkarev, Gipoteza J.Edmondsa i problema S.A.Kuka, Vestnik TGGPU, (24), 2, 2011 pp. 23-24.

11. S.V.Yablonskij, Vvedenie v diskretnuju matematiku, izd.Nauka,384,1986.

\end{document}